\documentclass[10pt,journal,compsoc]{IEEEtran}
\makeatletter
\newcommand\notsotiny{\@setfontsize\notsotiny{5.7}{8}}
\makeatother
\usepackage{amsmath,amsfonts}
\usepackage{algorithmic}
\usepackage{algorithm}
\usepackage{array}
\usepackage[caption=false,font=normalsize,labelfont=sf,textfont=sf]{subfig}
\usepackage{textcomp}
\usepackage{stfloats}
\usepackage{url}
\usepackage{verbatim}
\usepackage{graphicx}
\usepackage{cite}

\usepackage[table]{xcolor}
\usepackage{pifont}
\newcommand{\cmark}{\text{\ding{51}}}
\newcommand{\xmark}{\text{\ding{55}}}
\usepackage{multirow}
\usepackage{soul}
\usepackage[many]{tcolorbox}    
\newtcolorbox{mybox}[1]{%
    tikznode boxed title,
    enhanced,
    arc=0mm,
    interior style={white},
    attach boxed title to top center= {yshift=-\tcboxedtitleheight/2},
    fonttitle=\bfseries,
    colbacktitle=white,coltitle=black,
    boxed title style={size=normal,colframe=white,boxrule=0pt},
    title={#1}}

\begin{document}

\title{A Snapshot of the Mental Health of Software
Professionals}

%
%
%
%

\author{Eduardo~Santana~de~Almeida, 
        Ingrid~Oliveira~de~Nunes,
        Raphael~Pereira~de~Oliveira,
        Michelle~Larissa~Luciano~Carvalho,
        André~Russowsky~Brunoni,
        Shiyue~Rong,
        and~Iftekhar~Ahmed
\IEEEcompsocitemizethanks{\IEEEcompsocthanksitem E. Almeida was with the Institute of Computing (IC-UFBA), Federal University of Bahia, Brazil.\protect\\
E-mail: esa@rise.com.br
\IEEEcompsocthanksitem I. Ahmed and Shiyue Rong are with University of California, Irvine (UCI).}
\thanks{Manuscript received ...; revised...}}


\markboth{IEEE TRANSACTIONS ON SOFTWARE ENGINEERING, VOL. X, NO. Y, MARCH 2023}%
{Shell \MakeLowercase{\textit{et al.}}: A Sample Article Using IEEEtran.cls for IEEE Journals}

\IEEEtitleabstractindextext{
\begin{abstract}
Mental health disorders affect a large number of people, leading to many lives being lost every year. These disorders affect struggling individuals and businesses whose productivity decreases due to days of lost work or lower employee performance. Recent studies provide alarming numbers of individuals who suffer from mental health disorders, \textit{e.g.}, depression and anxiety, in particular contexts, such as academia. In the context of the software industry, there are limited studies that aim to understand the presence of mental health disorders and the characteristics of jobs in this context that can be triggers for the deterioration of the mental health of software professionals. In this paper, we present the results of a survey with 500 software professionals. We investigate different aspects of their mental health and the characteristics of their work to identify possible triggers of mental health deterioration. Our results provide the first evidence that mental health is a critical issue to be addressed in the software industry, as well as raise the direction of changes that can be done in this context to improve the mental health of software professionals.
\end{abstract}

}

\maketitle

\section{Introduction}\label{introduction}
On the 28th of July, Simone Biles, considered the greatest gymnast of all time, shocked the world after she withdrew from the women's individual all-around final at the Tokyo Olympic Games to focus on her mental health\footnote{https://edition.cnn.com/2021/07/28/sport/simone-biles-gymnastics-tokyo-2020-mental-health-spt-intl/index.html}. Simone Biles attitude is not an isolated case. According to a published report from American Psychological Association (APA) with emergent trends in psychology for 2021~\cite{Spiner:APA2021:2021Trends}, the national mental health crisis is among the top 10.

This crisis has shown unsettling aspects: two-thirds of employees report that poor mental health has undercut their job performance during the COVID-19 pandemic, and 40\% of employees are battling burnout. In addition, more than a third of Americans experienced clinical anxiety or depression symptoms. Employees' mental health struggles have an outsize impact on U.S. businesses also, with mental illness the leading cause of disability in the country, accounting for some 217 million days of lost work annually \cite{LyraHealth:guide2020:MentalHealthWorker}.

The mental health crisis has affected many sectors of society such as sports, entertainment \cite{Wilkes:misc2020:MentalHealthCinema}, health care \cite{Muller:PsyResearch2020:MentalHealthCOVID}, and information technology (IT) is not an exception since the software is literally "eating the world''~\cite{Andreessen:WSJ2011:SWImpact}. Based on a study with Indian professionals, Nayak~\cite{Nayak:IJHSSI2014:AnxietySE} identified that software developers have a considerably higher chance of experiencing fatigue, burnout, anxiety, and stress, compared to colleagues who perform mechanical tasks. 

In Stack Overflow's 2022 survey~\cite{StackOverflow:misc2022:DevSurvey}, which included participation from nearly 70k developers, respondents admitted having some type of mental issue, with memory disorder (10.6\%), anxiety (10.3\%) and depression (9.7\%) being the most common ones. These results indicate a significant prevalence of mental issues among software professionals.

Existing research usually focuses on a diversity of aspects related to the job of software engineers \cite{Li:ICSE2015:GreatSE, Li:ESE2020:GreatSE}, their mental model \cite{Latoza:ICSE2006:MentalModels} and work habits \cite{Claes:ICSE2018:OoOProgWork, MurphyHill:ICSE2019:ToolsToilet, Barton:TSE2020:MusicSE, 8823032, 9793986}, work conditions \cite{Meyer:TSE2017:PerceivedProductivity, Johnson:TSE2021:SEWorkEnvironment}, the need of sleep \cite{Fucci:TSE2018Sleep}, what makes a good day \cite{Meyer:TSE2021:SEGoodDay}, and their main motivations and satisfactions \cite{Graziotin:JSS2018:EffectsSEHappiness, 8851296}. 

Nevertheless, to the best of our knowledge, there are no studies reporting on the main aspects that affect the mental health of software professionals at work. Despite some reports \cite{Fujigaki:WDU1990:MentalWorkload, Nayak:IJHSSI2014:AnxietySE} of the occurrence of mental disorders such as depression and anxiety, very little is known about its prevalence among software professionals. In addition, human and social factors that directly impact the mental health of these individuals are also under-investigated.

This work aims to report the research findings of an investigation aimed at identifying the dimension of software professionals that suffer from mental disorders and the working context in which these individuals are more susceptible to developing them. We considered the working context into four directions: \textit{(i)} the work of software professionals concerning position, role, and  experience; \textit{(ii)} work environment; \textit{(iii)} schedule; and \textit{(iv)} any tasks undertaken. However, it was also important to consider the mental health history of software professionals and the impact caused by the COVID-19 pandemic \cite{DBLP:journals/ese/RalphBATKKNYDTZ20, DBLP:conf/icse/MillerRSF021, 9453143, DBLP:journals/tosem/FordSZBJMBHN22}.

Thus, to obtain a broad understanding of mental health issues among software professionals, we investigated the following research questions (RQs):

\textbf{RQ1.} \textit{What are the \textbf{mental health disorders} that affect software professionals, and what are their experiences associated with mental health?} 

\textbf{RQ2.} \textit{What \textbf{leisure and entertainment activities} do software professionals engage?} 

\textbf{RQ3.} \textit{What are the characteristics in the \textbf{work nature} that are associated with triggers for mental health disorders?}

\textbf{RQ4.} \textit{What are the characteristics in the \textbf{work environment} that might be triggers for mental health disorders?}

\textbf{RQ5.} \textit{What are the characteristics in the \textbf{work schedule} that might be triggers for mental health disorders?}

In RQ1, we aimed to investigate the mental health disorders that are primarily present among software professionals. In addition, even when mental health disorders are not present, there might be indicators that professionals are concerned with their mental health, such as by doing psychotherapy or reading on the topic. This is also investigated in this research question. RQ2 is used to observe external factors that may affect the mental health of our study participants, such as the frequency in which they engage social activity or practice sports. This is relevant because this can be a confounding factor in the job-related triggers of mental health disorders. These triggers are then investigated in RQ3-5, which aimed to identify triggers related to the work nature (\textit{e.g.}, type of work contract and position), work environment (\textit{e.g.}, relationship with colleagues, and pressure to stay up-to-date), and work schedule (\textit{e.g.}, amount of work hours, and unrealistic deadlines). 

To answer our research questions, we surveyed 500 software professionals from 35 countries. The survey asked respondents to provide feedback on their mental health, work environment characteristics, and daily activities. 

In summary, the main contributions of this paper are described as follows:

\begin{itemize}
    \item We collected a set of evidence that show how software development work can affect the mental health of software professionals;
    
    \item We also report the dimension of software professionals that suffer from mental health disorders beyond the depression and/or anxiety levels of these professionals; and
    
    \item We provide information on how organizations can offer better working conditions. In addition, we identified which and how leisure activities can be used as preventive measures beyond contributing to the well-being of software engineers.
\end{itemize}

\section{Background and Related Work}\label{related}

According to World Health Organization (WHO)\footnote{https://www.who.int/news-room/fact-sheets/detail/mental-health-strengthening-our-response}, Mental health is a state of well-being in which an individual realizes his or her abilities,  can cope with the normal stresses of life, can work productively and can contribute to his or her community.

Multiple social, psychological, and biological factors determine a person's mental health level at any time. For example, poor mental health is also associated with stressful work conditions, unhealthy lifestyles, and physical ill-health.


Despite the increasing importance of mental health, there is evidence that individuals have been struggling with mental disorders in many contexts. A recent study by \cite{Evans:NatureBio2018:MHGraduateEdu} reported alarming numbers of incidence of \emph{depression} and \emph{anxiety} among graduate students. Anecdotal evidence also shows that these mental disorders are also present among academics\cite{Wooldton:Nature2018:MentalHealth}.

According to the manual of mental disorders of the American Psychiatric Association\cite{APA:book2013:DSM-5}, depressed individuals exhibit sad, empty, or irritable moods. Their ability to function is largely affected by somatic and cognitive changes. Anxiety, in turn, refers to disorders that share features of excessive fear and anxiety and related behavioral disturbances. As opposed to fear, which is an emotional response to a real or perceived imminent threat, anxiety is the anticipation of a future threat. Among the mental disorders, the most impairing, common, and costly is major depression. Previous studies indicate that circa 5--12\% of males and 9--26\% of females have at least one episode of depression over the course of their life\cite{Crown:ClinicalPsy2002:Depression, Kessler:JAMA2003:Depression}.

Despite the increasing importance that mental health has received in other contexts, the software engineering community has only recently paid attention to aspects that can affect the mental health of software professionals. In this section, we present the main work discussing these aspects.  

\subsection{Depression and Anxiety in Software Development}

The software engineering field encompasses many social and human aspects \cite{Mayer:CSUR1981:PsychologyNovicesLearn, Franca:TSE2018:SEMotivation} that go beyond the technical aspects \cite{DeMarco:book2013:Peopleware}. These factors pressure the professionals involved in the software development process in such a way that generates a high level of depression, and anxiety \cite{Nayak:IJHSSI2014:AnxietySE, Franca:TSE2018:SEMotivation}. In general, software development is a highly collaborative process that includes large projects in terms of source code, developers, and users. In this sense, a major challenge facing software professionals is dealing not only with the high complexity of the technical aspects but also the human and social aspects that affect their performance at work \cite{Johnson:TSE2021:SEWorkEnvironment, Franca:TSE2018:SEMotivation}.

Psychotherapists state that software professionals often demonstrate feelings such as inadequacy, insecurity, low self-esteem, and dissatisfaction, resulting in problems in the social, marital, and sexual areas \cite{Nayak:IJHSSI2014:AnxietySE}. They also claim that these symptoms can manifest due to the work environment stemming from the growing demand for software that exceeds the capacity of software professionals. In addition, technology is changing fast, which makes it difficult for software professionals to keep updated at the same time as doing daily tasks at work \cite{Rajeswari:SIGMISCPR2003:StressSE}.

By interviewing systems analysts, Cohen~\cite{Cohen:HCI1984:ComputerUse} observed psychological problems such as irritability, depression, tension, and severe fatigue. 

To examine the relationships between job stressors and depressive symptoms, Haratani et al. \cite{Haratani1995} surveyed software engineers and managers in Japan. Engineers had significantly higher depression scores than managers. After controlling for confounding variables, interpersonal conflict in the project team and lack of control were common significant stressors for each group. In managers, job overload and changes in computer technology are significantly associated with depressive symptoms. 

Fujigaki~\cite{Fujigaki:IndustrialHealth1996:DepressionSE} conducted an observational study with ten male software engineers observing them every two weeks for five months. His goal was to conduct an observation of a time series of job-event/life events and depressive symptoms. The job events that were scored were the presence of time pressure/deadline, work overload, amount of work increase, responsibility increase, and trouble with clients. 

Rajeswari and Anantharaman \cite{Rajeswari:SIGMISCPR2003:StressSE} surveyed 156 software professionals from India to investigate factors that cause negative pressure on professionals. Ten key factors were identified. Fear of obsolescence and individual team interaction are two factors that explained twenty-five percent of the variance in the study, followed by the eight other factors, namely client interactions, work-family interface, role overload, work culture, technical constraints, family support to career, workload, and technical risk propensity. 

Rocha and Debert-Ribeiro \cite{Rocha:OEM2004:MeantalHealthSysAnalysis} carried out a cross-sectional study in two Brazilian companies with 636 system analysts to evaluate possible associations between working conditions and visual fatigue, mental and psycho-social health of systems analysts. They found that mental and psycho-social health aspects were associated with mental workload, difficulties with clients, strict deadlines, inadequate equipment, and work environment.

Nayak \cite{Nayak:IJHSSI2014:AnxietySE} surveyed 50 mechanical engineers and 50 software engineers from India to understand the dimensions contributing to their well-being. The study identified that software professionals experience a high level of anxiety than mechanical professionals. On the other hand, no significant difference was found between software and mechanical professionals related to the psychological dimension of mental health.

Despite studies focusing on the consequences of professional practice in the software development area, there is a significant research gap  related to the main aspects that affect the mental health of software professionals in the work context, mainly considering different countries and companies. In this sense, we conducted an investigation to collect evidence that contributes to understanding how software development work affects the mental health of professionals, especially considering aspects such as job characteristics, work environment, schedules, and tasks undertaken.

\section{Survey Design and Methodology}\label{survey}

To understand different aspects related to the mental health of software professionals, we conducted an online survey with 500 professionals. We designed our study as  a cross-sectional survey \cite{punch2003survey}, which involves collecting information from participants at one point in time. 

\subsection{Survey Construction}

Based on our research questions, we created a 20-minute survey questionnaire composed of nine parts: (1) \emph{Consent}: free and clarified consent term and consent to participate in the study; (2) \emph{Personal Data}: questions related to the demographic characteristics of the participants; (3) \emph{Mental Health History}: questions related to the participant mental health; (4) \emph{Leisure Activities}: questions related to how the participants enjoy their free time; (5) \emph{Job}: questions related to the characteristics of the nature of the job of participants (such as their position, team, and stability); (6) \emph{Work Environment}: questions related to the setup of the work environment; (7) \emph{Schedule}: questions related to the time available for the participants to complete their task, pressure, and workload (in terms of time); (8) \emph{Tasks}: questions related to the tasks performed by participants, such as whether they perceive value in what they do and the need for staying up-to-date; and finally; (9) \emph{Patient Health Questionnaire}: questions of a validated instrument to asses the current participants mental health. In the end, participants could also provide additional comments and provide their email in case they allowed us to contact them further or would like to receive the research results.

For selecting the questions to be included in the questionnaire related to software engineering, we used the demographic data typically collected in software engineering surveys and related studies, such as that by Johnson et al.~\cite{Johnson:TSE2021:SEWorkEnvironment}, who investigated the impact of the work environment on productivity and satisfaction. In addition, the authors that work on software engineering (and have experience as researchers and software engineers) first had a brainstorming session to identify possible questions. Then, they individually listed additional possible questions related to each topic. Next, all questions were assessed and possibly improved, discarded, or merged. 

One of the authors, who is a psychiatrist, reviewed and provided feedback on the questionnaire. In addition, he selected possible validated instruments for our patient health questionnaire. Two instruments were considered: (i) CIS-R~\cite{Brugha:PsyMedicine1999:CISR}, which diagnostic interview used for the National Survey of Psychiatric Morbidity in Great Britain; and (ii) PHQ-9~\cite{Kroenke:GenIntMedicine2001:PHQ9}, which is a reliable and valid measure of depression severity. While CIS-R, as opposed to PHQ-9, provides a diagnosis, it is long. Therefore, it could prevent participants from participating in the study. Both were assessed, and based on our pilot study (discussed as follows), PHQ-9 was selected. Due to space restrictions, we do not detail all questions in our questionnaire, which is available online.\footnote{Survey Instrument. https://github.com/MentalHealthSE/survey} 

\subsection{Ethical Considerations}
As there is a stigma and mental health is considered a taboo where people struggling with mental health can feel different than other people or feel like no one else understands, before sharing our survey, we submitted it to our IRB to minimize legal risks and ensure participants felt comfortable answering honestly. Thus, first, we made our survey anonymous and confidential: we did not collect names or IP addresses. Next, all participants gave informed consent. Third, we collected emails only from participants who would like to receive the study results. Finally, we are unable to publish our full dataset since it contains (although anonymous) demographic data that can inadvertently identify participants. 

\subsection{Survey Distribution}
\textbf{Survey platform.} Because of its simplicity and ease of use, we chose Google Forms to manage our survey. We scoped our study of mental health to software professionals around the world. All recruitment materials stated that the survey was optional, anonymous, and confidential. To mitigate participant self-selection bias, we stated that prior mental health issue was not required to participate. All data were collected between May-July 2021.

\textbf{Survey Pilot.} We piloted our survey with two software professionals from the industry to get feedback on the questions and their corresponding answers, the difficulties faced in answering the survey, and the time to finish it. As these pilot respondents were experts in the area, we also would like to know if we were asking the right questions. Next, we conducted another pilot study with two software developers (one from the open-source community and one from the industry). We conducted several iterations of the survey, rephrasing some questions and removing others to make the survey easier to understand and answer. Another concern in this stage was ensuring that the participants could finish it in 20-minutes. The pilot survey responses were used solely to improve the questions, and these responses were not included in the final results.

\textbf{Survey recruitment.} We followed a three-step approach to recruit survey respondents: initially, we posted survey information on personal accounts on social media (Twitter, LinkedIn, and Instagram). Next, the authors contacted potential respondents by email (convenience sampling) and asked them to share it with other potential respondents (snowballing). Because of this process, we could not track the total number of invitations.

\textbf{Survey data validation.} In total, we received 503 responses. We disqualified three responses: two responses that refused to participate and a repeated one, leading to 500 valid responses.

\textbf{Statistical Analysis.} After collecting the survey data, we used descriptive statistics and statistical tests to perform the analysis. 
For testing normality, we applied the Shapiro–Wilk Test~\cite{SHAPIRO1965}. In our analysis, Shapiro-Wilk Test showed a non normal distribution (\textit{p-value} less than 0.05). Thus, we applied the Wilcoxon Test \cite{woolson2007wilcoxon} to identify significant statistical difference between samples and the Cohen's d test to measure the effect size \cite{Kitchenham2017}. Finally, to identify a correlation between variables, we applied Spearman's test \cite{spearman1961proof}.

\section{Population Contextualization}\label{population}


Table \ref{tab:surveyParticipants} overviews our study population. The ages of participants ranged from 18 to 69, with an average age of 31.8. The majority (82\%) of our population are men which was similar to other studies in software engineering \cite{9793986}. The participants spread out in 35 countries across four continents. The top three countries where the respondents come from are Brazil, Germany, and United States. 

54\% of participants have a Bachelor's degree and 32\% have an advanced degree (Master/Ph.D.). The professional experience of these 500 respondents varies from no experience to 45 years, with an average of 9.2 years. The number of different companies which they worked ranged from 0 to 20 with an average of 3.8 companies. 

Regarding relationship status, there is a balance: 31\% are married and 30.2\% single. Only 26.2\% of participants are parents. 13\% of parents live with one children and 12.4\% live with two or three children. 

\begin{table*}

    \caption{Demographic characteristics of the participants (number and percentage, N = 500).}
    \label{tab:surveyParticipants}
    \centering
    \notsotiny
    \begin{tabular}{l l r l r l r l r l r}
        
        \textbf{Age (Years)}
            & 18--25 & 115 (23\%) & 26--35 & 238 (48\%) & 36--45 & 115 (23\%) & 46--55 & 26 (5\%) & 55+ & 6 (1\%) \\ 
        \textbf{Gender}
            & Female & 84 (16.8\%) & Male & 410 (82\%) & NA & 4 (0.8\%) & Other & 2 (0.4\%) \\ 
        \textbf{Country of Res.}        
            & Brazil & 374 (74.8\%) & Germany & 17 (3.4\%) & USA & 17 (3.4\%) & France  & 16 (3.2\%) & Other & 76 (15.2\% )\\ 
        \textbf{Education}
            & Less than & 2 (0.4\%) & Graduated & 46 (9\%) & Bachelor's & 270 (54\%) & Trade/technical & 21 (4.2\%) & Advanced & 161 (32.2\%) \\
            & high school &  & high school &  & degree &  & school &  & degree &  \\ 
        \textbf{Exp. (Years)}
            & 0--2 & 85 (17\%) & 3--5 & 123 (24.6\%) & 6--10 & 115 (23\%) & 10--15 & 87 (17.4\%) & 15+ & 90 (18\%) \\ 
        \textbf{Exp. (Companies)}       
            & 0--2 & 168 (33.6\%) & 3--5 & 236 (47.2\%) & 6--8 & 69 (13.8\%) & 9--13 & 20 (4\%) & 13+ & 7 (1.4\%) \\ 
        \textbf{Rel. Status}
            & Married & 155 (31\%) & Single & 151 (30.2\%) & In a relationship & 104 (20.8\%) & In a civil union & 44 (8.8\%) & Other & 46 (9.2\%) \\ 
        \textbf{Parent}
            & No & 369 (73.8\%) & Yes & 131 (26.2\%) \\ 
        \textbf{Children}
            & 0 & 373 (74.6\%) & 1 & 65 (13\%) & 2--3 & 62 (12.4\%) \\
    \end{tabular}
\end{table*}

\section{Survey Results}\label{results}


This section describes the survey results. We organize the survey analysis around our research questions:






\subsection{RQ1: Mental Health of Software Professionals}

\begin{figure}[b]
    \centering
    \includegraphics[width=\columnwidth]{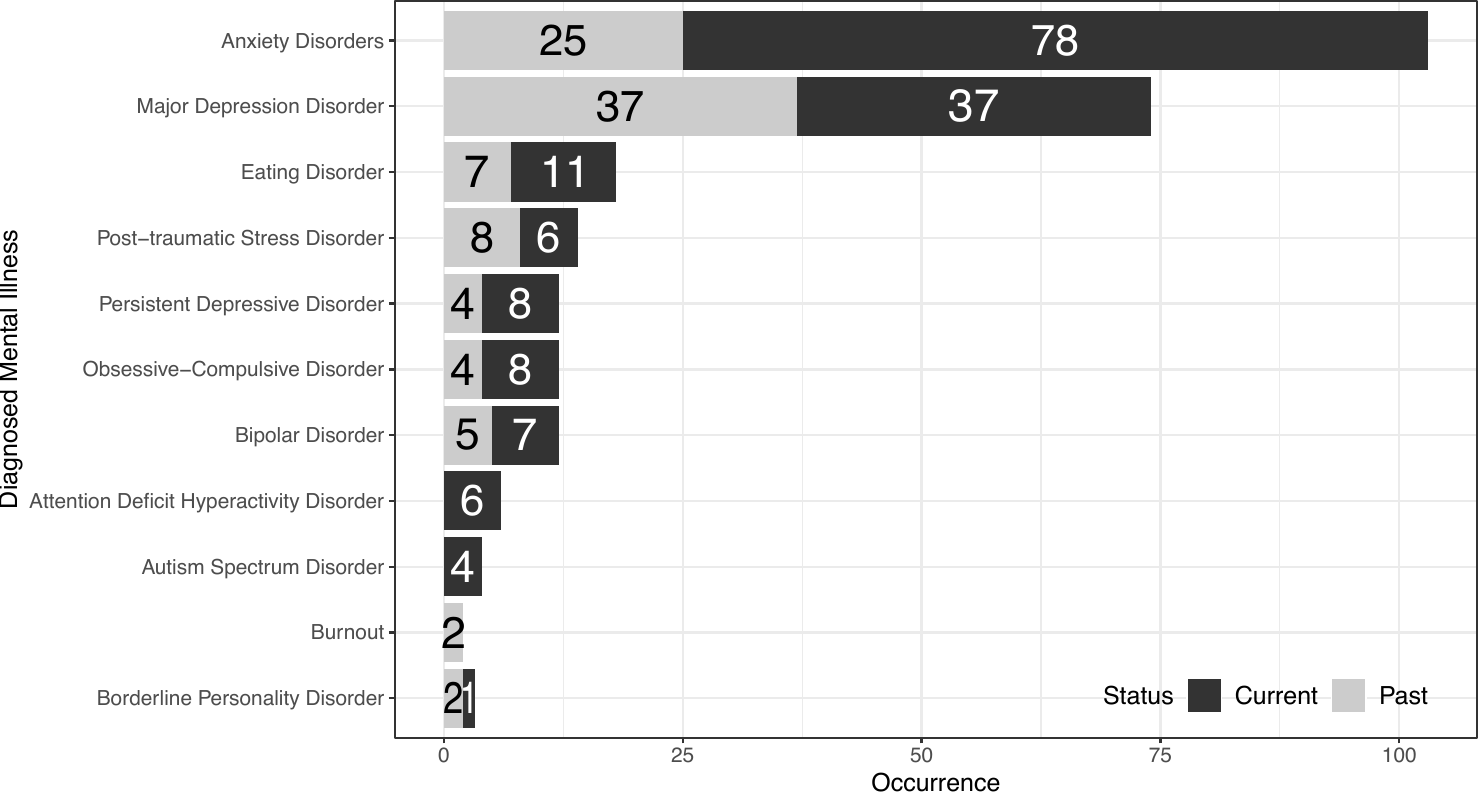}
    \caption{Mental Health Disorders among Software Professionals.}
    \label{fig:RQ1-2}
\end{figure}

Overall, we find that 30.2\% (151/500) of our participants were diagnosed with mental health disorders in the past or currently. As it can be seen in Figure~\ref{fig:RQ1-2}, anxiety disorders and depression are representative: 103 participants had anxiety (20.6\%) and 74 (14.8\%) had depression.

It is important to highlight that these mental disorders are not isolated cases in some situations. 65 participants reported more than one mental disorder, and one participant reported being diagnosed with five mental disorders. The participants (P) also highlighted the impact on their work and relationship with co-workers: 

\xmark \textit{P 282: "Wanted to end my life. Ended up loosing my job instead.}"

\xmark \textit{P 39: "Fear of meetings with my boss, sleep deprivation, and permanent stress. I had to take medicine.}"

\xmark \textit{P 101: "Fell tired all the time. No disposition to social interaction.}"

\vspace{-0.5cm} 
\begin{mybox}{Observation 1}
\small
30.2\% (151/500) of our participants were diagnosed with mental health disorders in the past or currently. Anxiety (20.6\%) and depression (14.8\%) were the most representative disorders.
\end{mybox}

\begin{figure}[b]
    \centering
    \includegraphics[width=\columnwidth]{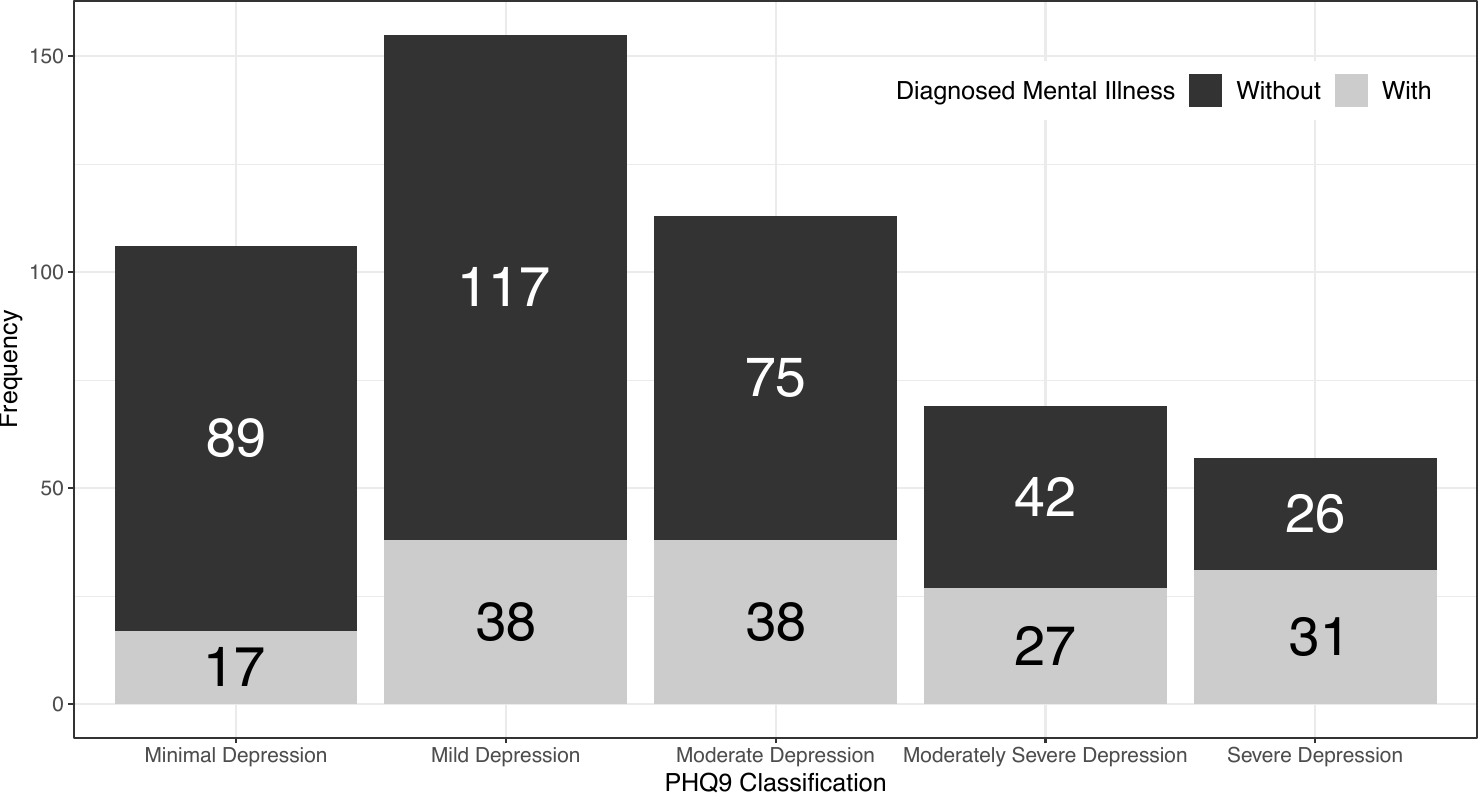}
    \caption{Depression Levels of Software Professionals according to PHQ-9  classification.}
    \label{fig:RQ1}
\end{figure}

\textbf{PHQ-9 Classification.} The PHQ-9 is a multipurpose instrument for screening, diagnosing, monitoring and measuring the severity of depression. It is composed of nine questions which analyze different aspects of the patient over the last 2 weeks. Figure~\ref{fig:RQ1} shows the distribution of the depression level of software professionals according to PHQ-9 classification. 

In our population, 126 participants (25.2\%) have moderately severe depression or severe depression. 69 participants (13.8\%) had moderately severe depression which generally warrants treatment for depression using medication, therapy, or a combination of the two and 57 participants (11.4\%) have severe depression which also warrants the same treatment. 

An important remark is that analyzing the participants without any record of mental health disorders (349 participants - 69.8\%), we found that according to PHQ-9 classification, 75 participants (21.5\%) had moderate depression, 42 participants had moderately severe depression (12\%) and 26 (7.4\%) participants had severe depression. 

\vspace{-0.3cm} 
\begin{mybox}{Observation 2}
\small
25.2\% (126/500) of our participants were diagnosed with moderately severe depression or severe depression, which generally warrants treatment for depression using medication, therapy, or a combination of the two. Moreover, 68 participants (19.4\%) without any record of mental health disorders were  diagnosed with moderately severe depression or severe depression.
\end{mybox}

\textbf{Mitigation strategies.} Software professionals have used different ways to deal with mental health disorders. 249 participants (50\%) have already done therapy from which 104 (21\%) currently do to mitigate the consequences of mental health disorders. 346 participants (69\%) have already searched the web for mental health information. YouTube, Medium and Reddit are the top 3 places to find information related to mental health. The participants (184 - 37\%) also read book, the majority of them (95 - 52\%) read 1-2 books related to mental health. An worrisome aspect is that 107 participants (21.4\%) have already taken drugs, 39 (8\%) without medical prescription.

\textbf{COVID-19 pandemic.} Different studies in software engineering \cite{DBLP:journals/ese/RalphBATKKNYDTZ20, DBLP:conf/icse/MillerRSF021, DBLP:journals/tosem/FordSZBJMBHN22, 9453143} have discussed the impact of COVID-19 in software development. In our study, 286 respondents (57\%) said that the pandemic period impacted negatively their mental health. 153 respondents were not affected and 35 respondents (7\%) said that during the pandemic their mental health improved.

\subsection{RQ2: Leisure and Entertainment Activities}
Different studies have indicated the long-term benefits of leisure and entertainment activities on mental health and well-being \cite{Goodman1, Goodman, TONIETTO2021104198} of unemployed and employed workers in general. We investigated which leisure and entertainment activities software professionals engage and how the COVID-19 pandemic impacted them, since during the pandemic different restrictions and lockdown were imposed around the world.

Before the pandemic, 388 participants (77.6\%) said they performed physical exercise (walking, running, swimming, and so on) sometimes (99 - 19.8\%), usually (144 - 28.8\%), or always (145 - 29\%). Regarding social activities, which involved meet friends and go out to restaurants and bars, 435 participants (87\%) reported they sometimes (162 - 32.4\%), usually (166 - 33.2\%), or always (107 - 21.4\%) engage them. 

445 participants (89\%) sometimes (155 - 31\%), usually (185 - 37\%), or always (105 - 21\%) dedicate some time to hobbies, such as cooking, reading, playing an instrument, and so on. Finally, we investigated also how the software professionals spend time with family without any concern related to the job activities. 404 participants (80.8\%) reported that they sometimes (118 - 23.6\%), usually (178 - 35.6\%), or always (108 - 25.6\%) spend time with family. 

\begin{figure*}
    \centering
    \includegraphics[width=1.0\textwidth]{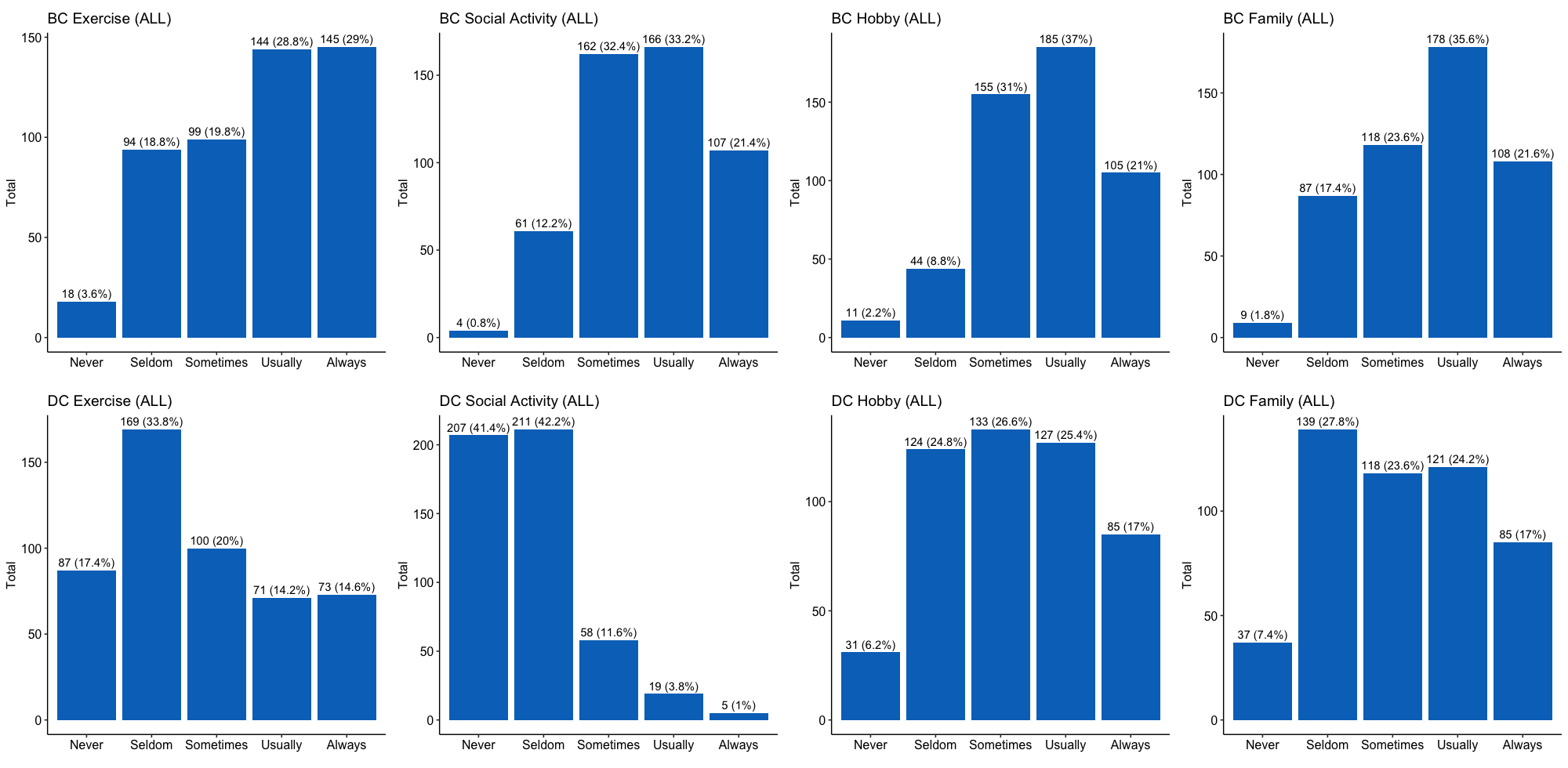}
    \caption{Summary of Leisure and Entertainment Activities before and during the pandemic. BC= Before Covid and DC= During Covid}
    \label{fig:RQ2}
\end{figure*}

Figure~\ref{fig:RQ2} shows how software professionals engaged in leisure and entertainment activities before and during the pandemic and 
Table \ref{tab:resultsActLeisure} reports the comparison statistics results.

\vspace{-0.4cm} 
\begin{table}[h]
\notsotiny
\centering
\caption{Before x During Activities and Leisure Results.}
\label{tab:resultsActLeisure}
\begin{tabular}{c|c|c|c}
\hline
      \textbf{Activity / Leisure}            & \textbf{Wilcoxon (\textit{p}-value)}  & \textbf{Cohen's d} &
      \textbf{In favor of}\\ \hline
Physical exercise & 2.2e-16   & 0.68 {[}0.56, 0.81{]} (medium) & Before\\ \hline
Social activities & 2.2e-16   & 1.97 {[}1.81, 2.12{]} (large) & Before \\ \hline
Hobbies           & 1.537e-09 & 0.40 {[}0.27, 0.52{]} (small) & Before \\ \hline
Time with family  & 2.567e-08 & 0.36 {[}0.24, 0.49{]} (small) & Before \\ \hline
\end{tabular}
\end{table}

\vspace{-0.4cm} 
\begin{mybox}{Observation 3}
\small
During the COVID-19 pandemic, software professionals reduced the amount of physical exercises, social activities, hobbies and engagement with family members. 
\end{mybox}

\textbf{No History x History.} Table \ref{tab:resultsActLeisureCMP} shows the significant statistics results when comparing the participants \textit{without history} of mental health disorders (349/500) and participants \textit{with history} of mental health disorders (151/500) before and during the pandemic.

\vspace{-0.6cm} 
\begin{table}[h]
\notsotiny
\centering
\caption{No History x History Results.}
\label{tab:resultsActLeisureCMP}
\begin{tabular}{c|c|c|c|c}
\hline
      \textbf{Period} &
      \textbf{Activity / Leisure}            & \textbf{Wilcoxon (\textit{p}-value)}  & \textbf{Cohen's d} &
      \textbf{In favor of}\\ \hline
\multirow{3}{*}{Before}& Physical exercise & 0.01086    & -0.22 [-0.41, -0.03] (small) & No history\\ \cline{2-5}
& Hobbies         & 0.02926     & -0.21 [-0.40, -0.02] (small) & No history \\ \cline{2-5}
& Time with family  & 1.133e-06         & -0.49 [-0.68, -0.30] (small)  & No history\\ \hline
During& Time with family & 0.005383     & -0.27 [-0.47, -0.08] (small) & No history \\ \hline
\end{tabular}
\end{table}




\vspace{-0.4cm} 
\begin{mybox}{Observation 4}
\small
Before the pandemic, software professionals \textit{with history} of mental health disorders did less physical exercise and had less hobbies. Moreover, before and during the pandemic this group also spent less time with their family. 
\end{mybox}

\subsection{RQ3: Work Nature}


Previous studies \cite{Perkbox} have identified that the level of occupational stress workers tend to feel directly correlates to the size of the business they are employed by. According to the study, micro businesses employing a maximum of four members of staff had the lowest instances of workers reporting job related stress of companies of any other size.

In our study, we investigated organization size and other aspects related to work nature which could be triggers for mental health disorders, such as the team size, the work type, and the employees role in a project. 

\textbf{Organization size.} From our population, we find that 42.2\% of respondents work for organizations with 1000 or more employees, 20.4\% of the participants work for organizations with 100-499 employees, 16.6\% of the respondents work for organizations with 10-99 employees, 9.2\% of the respondents work for organizations with 500-999 employees, and 5.6\% of the participants work for organizations with 0-9 employees. 4.6\% and 1.4\% of respondents, respectively, did not know or were not applicable.

Analyzing the participants with history of mental health disorders (151/500) and each organization size (0-9, 10-99, 100-499, 500-999, 1000 or more, Do not know), we did not find any correlation after applying the Spearman's test (all the \textit{p}-values were $\geq$ 0.05).

\textbf{Team size.} Different studies in software engineering have discussed the impact of team size on software development effort \cite{10.5555/1324991.1324992}, cost \cite{10.1145/1435417.1435449}, and productivity \cite{10.1016/j.jss.2011.09.009}. While Scrum methodology \cite{10.5555/2380978} and companies like Amazon (Jeff Bezos's two-pizza rule\footnote{https://docs.aws.amazon.com/whitepapers/latest/introduction-devops-aws/two-pizza-teams.html}) advocate the adoption of small teams for better communication and collaboration, no study has investigated the impact of team size on mental health. 

From our survey, we found that 71\% of respondents work in teams with 1-10 members, 15.6\% of the participants work in teams with 11-20 members, 3.6\% of respondents work in teams with 31 or more members, 2.8\% of the participants work in teams 21-30 members, and 7\% was not applicable. After applying the Spearman's test to identify a correlation between the participants with history of mental health disorders (151/500) and each one of the team sizes (1-10, 11-20, 21-30, 31 or more), we did not find any significant correlation (all the \textit{p}-values from Spearman's test were $\geq$ 0.05).

\textbf{Work type.} As software professionals can work in different ways (freelancer, open-source project, private organization, and so on), we investigated whether specific work types could impact in some way their mental health. 92.7\% (464/500) of our participants work for an organization, 13\% of participants contribute to open-source projects, 11.4\% of participants work as freelancer, 1.8\% of participants were unemployment, and 4.6\% were not applicable.         

Considering the participants with history of mental health disorders (151/500) and each one of the work types (\textit{organization}, \textit{freelancer}, \textit{OSS}, \textit{unemployed}, \textit{other}), 
we applied the Spearman's test and we found a significant correlation between \textit{organization} and participants with mental health disorders
(\textit{p}-value = 0.02). However, the correlation was negative (\textit{rho} = -0.10), implying that if the number of participants working in a organization increase, the number of participants with history of mental health disorders decrease. Regarding \textit{freelancer}, we found a significant (\textit{p}-value = 0.03) and positive (\textit{rho} = 0.09) correlation with participants with mental health disorders. Thus, we found that \textit{freelancer} is a trigger to mental health disorders. 

\textbf{Roles.} Previous research in psychology \cite{Kelloway, pub.1029706050} has suggested over the decades that job characteristics and role stressors are related to the measures of job-related mental health, such as work dissatisfaction, burnout, and depression. Role stress is the stress experienced by workers because of their role (job) in the organization. In a software development project, software professionals act in different roles with specific activities, responsibilities and concerns. In our survey, we investigated the possible impact of roles on mental health. Unlike other areas \cite {doi:10.1177/002224298404800402, 10.1046/j.1442-2018.2001.00086.x, 10.1111/j.1442-2018.2005.00221.x, ijerph16132394} few is known about this aspect in software development.   

78.2\% (391/500) of our respondents work as software engineers, 20.4\% of respondents work as software architects, 14.4\% of respondents work as project managers, 9.6\% of respondents work as requirements analysts, 7.6\% of respondents work as test engineers, and 10.8\% of respondents work in other roles, such as data scientist, head of quality assurance, and so on. 

After applying the Spearman's test to identify correlations between the participants with history of mental health disorders (151/500) and each one of the roles (\textit{Product Manager}, \textit{Requirements Analyst}, \textit{Software Architect}, \textit{Software Engineer/Programmer}, \textit{Test Engineer}, \textit{Other}), we did not find any significant correlation for \textit{Product Manager}, \textit{Requirements Analyst}, \textit{Software Architect}, \textit{Software Engineer/Programmer}, and \textit{Other} (all p-values were $\geq$ 0.05). However, we found a significant (\textit{p}-value = 0.0004) and positive (\textit{rho} = 0.15) correlation after applying the Spearman's test to check the correlation of participants with history of mental health disorders and \textit{Test Engineer}. Hence, we found that the role of \textit{Test Engineer} is associated with mental health disorders.

\vspace{-0.3cm} 
\begin{mybox}{Observation 5}
\small
Analyzing software professionals with history of mental health disorders, we found that the work type (freelancer) and role (test engineer) are associated with mental health disorders. 
\end{mybox}

\subsection{RQ4: Work Environment}

The connection between working conditions and mental health has been long conceptualized, discussed, and documented in the epidemiological, medical, and sociological literature \cite{Harvey301, 10.1016/j.labeco.2022.102176}. The Mental Health Foundation (2021), for example, has recently recognized that working conditions and the environment can significantly affect mental health \cite{10.1016/j.labeco.2022.102176}. Recently, the software engineering area has also starting paying attention to effect of work environments on productivity and satisfaction of software engineers \cite{Johnson:TSE2021:SEWorkEnvironment}. However, aspects related to mental health are not explored.

In our survey, we investigated which characteristics of the work environment (physical structure, leisure facilities, resting facilities, social activities, relationship with colleagues, and dissatisfaction with the job) can be triggers for mental health disorders.

\textbf{Physical work environment.} From our population, we find that before COVID-19, 42.4\% of our participants worked in private workspaces, 37.2\% of participants worked in open workspaces, 15.6\% of participants worked in home office, and 4.8\% of participants worked in other work environments, such as public spaces and coffee shops. The following are some feedback from participants' satisfaction regarding their physical work environment before COVID-19:

\cmark \textit{P 64: "Frequent contact with different co-workers from different teams who helped in aspects related to innovation, motivation, and understanding of general goals of the organization.}"

\cmark \textit{P 14: "Perfect place, nice landscape, infrastructure with places for drinking coffee and chat with colleagues, meeting rooms, video games, and good lighting.}"

\xmark \textit{P 1: "Noise from parallel conversation, interruptions, and fix working hour.}"

\xmark \textit{P 28: "Too many people. Uncomfortable chairs and tables. Long commute.}"

During the COVID-19 pandemic, 94.2\% of our respondents worked in home office, 2.2\% of respondents worked in private workspaces, 0.8\% of respondents worked in open spaces, and 2.8 \% of participants worked in hybrid environments (home office/private workspace, home office, co-working spaces). 

On the overall satisfaction with the work environment, 68.8\% of the survey respondents are satisfied with the working environment. 98, 48, and 9 respondents are neutral, dissatisfied, and very dissatisfied, respectively. The average Likert score for this statement is 3.79 (\textit{i.e.}, between ``Neutral'' and ``Satisfied'').

\vspace{-0.3cm} 
\begin{mybox}{Observation 6}
\small
68.8\% of the survey respondents are satisfied with the working environment. 
\end{mybox}

\textbf{Leisure and Resting facilities.} 341 participants work in organizations which provide leisure facilities (gym or game room). Nevertheless, 61\% of these participants never used the leisure facilities. 62, 39, and 26 participants used rarely, sometimes, and usually, respectively. The average Likert score for this statement is 1.71 (\textit{i.e.}, between ``Never'' and ``Rarely'').

Regarding resting facilities, 340 respondents work in organizations which provide resting facilities, such as massage or nap room. However, 60.88\% of these respondents never used the resting facilities. 61, 40, and 19 respondents used rarely, sometimes, and usually, respectively. The average Likert score for this statement is 1.74 (\textit{i.e.}, between ``Never'' and ``Rarely'').

\vspace{-0.3cm} 
\begin{mybox}{Observation 7}
\small
61\% of the survey respondents never used the leisure facilities in their organizations. Regarding resting facilities, 60.88\% of respondents never used them. 
\end{mybox}

\textbf{Social Activities.} 412 participants work in organizations which organize social activities for their employees. However, 26.70\% of these participants never participate in social activities. 73, 66, and 73 participants participated rarely, sometimes, and usually, respectively. The average Likert score for this statement is 2.90 (\textit{i.e.}, between ``Rarely'' and ``Sometimes'').

\vspace{-0.3cm}
\begin{mybox}{Observation 8}
\small
26.70\% of the survey respondents never participated in social activities organized by their company. 
\end{mybox}

\textbf{Relationship with Colleagues.} We asked participants if they have a healthy relationship with their colleagues. Overall, 87\% of the participants have a healthy relationship with their colleagues. 35, 8, and 3 respondents are neutral, disagree, and strongly disagree, respectively. The average Likert score for this statement is 4.40 (\textit{i.e.}, between ``Agree'' and ``Strongly Agree'').

\vspace{-0.3cm} 
\begin{mybox}{Observation 9}
\small
The majority of the survey respondents (87\%) have a healthy relationship with their colleagues. 
\end{mybox}

\textbf{Dissatisfaction with the Job.} We asked participants if they are always dissatisfied with their job. Overall, 35\% of the participants disagreed with this statement. 123, 136, and 52 respondents are neutral, agree, and strongly agree, respectively. The average Likert score for this statement is 3.01 (\textit{i.e.}, between ``Neutral'' and ``Agree'').

\vspace{-0.2cm} 
\begin{mybox}{Observation 10}
\small
Several survey respondents (38\%) are dissatisfied with their job. 
\end{mybox}



After applying Spearman's test to correlate each one of the levels of work environment characteristics with participants with history of mental health disorders, we found five significant correlations shown in Table \ref{table:workEnvironmentAnalysis}.
The other tests (available online\footnote{Survey Instrument. https://github.com/MentalHealthSE/survey})
did not present a significant correlation. 


\begin{mybox}{Observation 11}
\small
Before COVID-19, working at home office represented a trigger for mental health disorders. Moreover, rarely frequency to social activities explicitly organized by the company represents a trigger as well. 
Also, survey respondents who are neutral or strongly agree with healthy relationships with colleagues, are less prone to have history of mental health disorders. Finally, survey respondents often dissatisfied with their job are more prone to have history of mental health disorders.
\end{mybox}

\begin{table}[h]
\scriptsize
\centering
\caption{Work Environment Correlation Results}
\label{table:workEnvironmentAnalysis}
\begin{tabular}{c|c|cc}
\hline
\multirow{2}{*}{\textbf{\begin{tabular}[c]{@{}c@{}}Working Environment\\ Characteristics\end{tabular}}} & \multirow{2}{*}{\textbf{\begin{tabular}[c]{@{}c@{}}Correlation with History\\ of Mental Health Disorders\end{tabular}}} & \multicolumn{2}{c}{\textbf{Spearman's Test}}                 \\ \cline{3-4} 
                                                                                                        &                                       & \multicolumn{1}{c|}{\textbf{\textit{p}-value}}                & \textbf{\textit{rho}}                         \\ \hline 
\multirow{4}{*}{\begin{tabular}[c]{@{}c@{}}Work\\ Environment\\ {[}before COVID-19{]}\end{tabular}}     & \cellcolor{gray!25}Home Office        & \multicolumn{1}{c|}{\cellcolor{gray!25}0.0049}        & \cellcolor{gray!25}0.1253             \\ \cline{2-4} 
                                                                                                        & Open Workspace                        & \multicolumn{1}{c|}{0.7133}                           & 0.0164                                \\ \cline{2-4} 
                                                                                                        & Private Workspace                     & \multicolumn{1}{c|}{0.0298}                           & -0.0971                               \\ \cline{2-4} 
                                                                                                        & Other                                 & \multicolumn{1}{c|}{0.5705}                           & -0.0254                               \\ \hline

\multirow{4}{*}{\begin{tabular}[c]{@{}c@{}}Work\\ Environment\\ {[}during COVID-19{]}\end{tabular}}     & Home Office                           & \multicolumn{1}{c|}{0.4648}                           & 0.0327                                 \\ \cline{2-4} 
                                                                                                        & Open Workspace                        & \multicolumn{1}{c|}{0.1873}                           & -0.0590                               \\ \cline{2-4} 
                                                                                                        & Private Workspace                     & \multicolumn{1}{c|}{0.3810}                           & -0.0392                               \\ \cline{2-4} 
                                                                                                        & Other                                 & \multicolumn{1}{c|}{0.6493}                           & 0.0203                                \\ \hline

\multirow{4}{*}{\begin{tabular}[c]{@{}c@{}}Satisfaction with the\\ Working Environment\end{tabular}}    & Very satisfied                        & \multicolumn{1}{c|}{0.2084}                           & -0.0563                                \\ \cline{2-4} 
                                                                                                        & Satisfied                             & \multicolumn{1}{c|}{0.1412}                           & 0.0659                                \\ \cline{2-4} 
                                                                                                        & Neutral                               & \multicolumn{1}{c|}{0.8840}                           & -0.0065                               \\ \cline{2-4} 
                                                                                                        & Dissatisfied                          & \multicolumn{1}{c|}{0.6217}                           & -0.0221                               \\ \cline{2-4}
                                                                                                        & Very dissatisfied                     & \multicolumn{1}{c|}{0.8367}                           & 0.0092                                \\ \hline

\multirow{4}{*}{\begin{tabular}[c]{@{}c@{}}How Often do you\\ use Leisure Facilities\end{tabular}}      & Never                                 & \multicolumn{1}{c|}{0.8154}                           & 0.0104                                 \\ \cline{2-4} 
                                                                                                        & Rarely                                & \multicolumn{1}{c|}{0.9352}                           & 0.0036                                \\ \cline{2-4} 
                                                                                                        & Sometimes                             & \multicolumn{1}{c|}{0.9359}                           & 0.0036                                \\ \cline{2-4} 
                                                                                                        & Usually                               & \multicolumn{1}{c|}{0.9484}                           & 0.0029                                \\ \cline{2-4}
                                                                                                        & Always or Almost Always               & \multicolumn{1}{c|}{0.4686}                           & -0.0324                               \\ \cline{2-4}
                                                                                                        & Company does not provide              & \multicolumn{1}{c|}{0.5489}                           & -0.0268                               \\ \hline 

\multirow{4}{*}{\begin{tabular}[c]{@{}c@{}}How Often do you\\ use Resting Facilities\end{tabular}}      & Never                                 & \multicolumn{1}{c|}{0.3730}                           & -0.0399                                \\ \cline{2-4} 
                                                                                                        & Rarely                                & \multicolumn{1}{c|}{0.0504}                           & 0.0875                                \\ \cline{2-4} 
                                                                                                        & Sometimes                             & \multicolumn{1}{c|}{0.6989}                           & -0.0173                               \\ \cline{2-4} 
                                                                                                        & Usually                               & \multicolumn{1}{c|}{0.1637}                           & -0.0623                               \\ \cline{2-4}
                                                                                                        & Always or Almost Always               & \multicolumn{1}{c|}{0.5718}                           & -0.0253                               \\ \cline{2-4}
                                                                                                        & Company does not provide              & \multicolumn{1}{c|}{0.5323}                           & 0.0279                                \\ \hline

\multirow{4}{*}{\begin{tabular}[c]{@{}c@{}}How Often do you\\ go to Social Activities\\ explicitly organized by\\ the company\end{tabular}}     & Never                                 & \multicolumn{1}{c|}{0.4500}                           & -0.0338                                \\ \cline{2-4} 
                                                                                                        & \cellcolor{gray!25}Rarely             & \multicolumn{1}{c|}{\cellcolor{gray!25}0.0282}     & \cellcolor{gray!25}0.0981                        \\ \cline{2-4} 
                                                                                                        & Sometimes                             & \multicolumn{1}{c|}{0.2426}                           & 0.0523                                \\ \cline{2-4} 
                                                                                                        & Usually                               & \multicolumn{1}{c|}{0.5734}                           & -0.0252                               \\ \cline{2-4}
                                                                                                        & Always or Almost Always               & \multicolumn{1}{c|}{0.1176}                           & -0.0700                               \\ \cline{2-4}
                                                                                                        & Company does not provide              & \multicolumn{1}{c|}{0.2991}                           & -0.0465                               \\ \hline

\multirow{4}{*}{\begin{tabular}[c]{@{}c@{}}I have a healthy\\ relationship with\\ my colleagues\end{tabular}}           & Strongly disagree     & \multicolumn{1}{c|}{0.4239}                           & -0.0358                                \\ \cline{2-4} 
                                                                                                        & Disagree                              & \multicolumn{1}{c|}{0.1613}                           & 0.0627                                \\ \cline{2-4} 
                                                                                                        & \cellcolor{gray!25}Neutral            & \multicolumn{1}{c|}{\cellcolor{gray!25}0.0485}                                & \cellcolor{gray!25}-0.0882                            \\ \cline{2-4} 
                                                                                                        & Agree                                 & \multicolumn{1}{c|}{0.3124}                           & 0.0452                                \\ \cline{2-4}
                                                                                                        & \cellcolor{gray!25}Strongly agree     & \multicolumn{1}{c|}{\cellcolor{gray!25}0.0227}                                & \cellcolor{gray!25}-0.1018                            \\ \hline

\multirow{4}{*}{\begin{tabular}[c]{@{}c@{}}I am often\\ dissatisfied with my job\end{tabular}}          & Strongly disagree                     & \multicolumn{1}{c|}{0.1845}                           & -0.0594                                \\ \cline{2-4} 
                                                                                                        & Disagree                              & \multicolumn{1}{c|}{0.0908}                           & -0.0756                               \\ \cline{2-4} 
                                                                                                        & Neutral                               & \multicolumn{1}{c|}{0.1690}                           & -0.0616                               \\ \cline{2-4} 
                                                                                                        & Agree                                 & \multicolumn{1}{c|}{0.5939}                           & 0.0239                                \\ \cline{2-4}
                                                                                                        & \cellcolor{gray!25} Strongly agree    & \multicolumn{1}{c|}{\cellcolor{gray!25}0.0324}        & \cellcolor{gray!25}0.0956                             \\ \hline 

\end{tabular}
\end{table}

\vspace{-0.4cm} 
\begin{table}[h]
\notsotiny
\centering
\caption{Work Environment Correlation Results}
\label{table:workEnvironmentAnalysis}
\begin{tabular}{c|c|cc}
\hline
\multirow{2}{*}{\textbf{\begin{tabular}[c]{@{}c@{}}Working Environment\\ Characteristics\end{tabular}}} & \multirow{2}{*}{\textbf{\begin{tabular}[c]{@{}c@{}}Correlation with History\\ of Mental Health Disorders\end{tabular}}} & \multicolumn{2}{c}{\textbf{Spearman's Test}}                 \\ \cline{3-4} 
                                                                                                        &                                       & \multicolumn{1}{c|}{\textbf{\textit{p}-value}}                & \textbf{\textit{rho}}                         \\ \hline 
\begin{tabular}[c]{@{}c@{}}Work Environment {[}before COVID-19{]}\end{tabular}     & Home Office        & \multicolumn{1}{c|}{0.0049}        & 0.1253             \\  \hline

\begin{tabular}[c]{@{}c@{}} Social Activities\end{tabular}      &  Rarely               & \multicolumn{1}{c|}{0.0282}     & 0.0981      \\ \hline

\multirow{2}{*}{\begin{tabular}[c]{@{}c@{}}Relationship with Colleagues\end{tabular}}           & Neutral               & \multicolumn{1}{c|}{0.0485}                           & -0.0882       \\ \cline{2-4} 
                                                                                                        & Strongly agree        & \multicolumn{1}{c|}{0.0227}                           & -0.1018       \\ \hline

\begin{tabular}[c]{@{}c@{}}Dissatisfaction with the Job\end{tabular}            &  Strongly agree       & \multicolumn{1}{c|}{0.0324}   & 0.0956                                \\ \hline 

\end{tabular}
\end{table}

\subsection{RQ5: Work Schedule}

Evidence has shown that several types of job stressors, including workplace conditions, can influence the onset and progress of mental health disorders \cite{10.1017/S0033291711000171}. One of the main conditions affecting workers' mental health is working hours. Some empirical research suggests a close relationship between working hours and workers' mental health \cite{10.1097/JOM.0b013e31829b27fa, 10.1136/jech-2018-211309, 10.1093/occmed/kqx054, ijerph16122102, 10.1016/j.socscimed.2019.112774}. Additionally to the number of working hours, other work schedule characteristics, such as the frequency of night work and short daily rest periods (quick return), can affect workers' health as work-related stressors \cite{10.1017/S0033291711000171}. In our survey, we investigated which six work schedule characteristics (working hours, work on weekends, checking messages outside working hours, unrealistic deadlines, the impact of COVID-19 pandemic, and tasks performed) can affect software professionals’ mental health.

\textbf{Working hours.} We asked the survey participants, the number of hours they are expected to work in a typical week and the actual number of worked hours. 
Regarding all participants, the median for both expected and actual worked hours is 40 hours. However, their means are different (38.37 hours for expected and  40.97 hours for actual). When we analyzed only the participants with history of mental health, we found that the median for both is also 40 hours and their means are different (38.66 hours for expected and  40.83 hours for actual).
Table \ref{tab:resultsWorkingHours} shows the statistics results after applying the Wilcoxon test to compare the difference between expected and actual worked hours.




\vspace{-0.5cm}
\begin{table}[h]
\notsotiny
\centering
\caption{Expected x Actual Working Hours Results.}
\label{tab:resultsWorkingHours}
\begin{tabular}{c|c|c|c}
\hline
      \textbf{Population}               & \textbf{Wilcoxon (\textit{p}-value)}  & \textbf{Cohen's d} &  \textbf{In favor of}\\ \hline
All                                     & 2.318e-08     & -0.27 [-0.39, -0.14] (small) & Actual \\ \hline
History of mental health                & 0.009         & -0.21 [-0.44, 0.01] (small) & Actual  \\ \hline
\end{tabular}
\end{table}

\vspace{-0.3cm}
\begin{mybox}{Observation 12}
\small
Software professionals work more hours than expected and it can be a trigger to deteriorate their mental health.
\end{mybox}

\textbf{Weekends.} We found that 166 out of 500 survey respondents never worked during the weekend. In addition, a substantial number of respondents worked rarely or sometimes (174 rarely respondents and 108 sometimes respondents). The average Likert score for this statement is 2.12 (\textit{i.e.}, between ``Rarely'' and ``Sometimes''). Considering only the participants with history of mental health, we did not find any significant correlation with each one of the weekend's possible answers after applying Spearman's test. 

\textbf{Messages.} In a typical workday, software professionals use different communication tools, such as emails, meeting notifications, real-time messaging and chat, and remote video calls. We asked the survey participants how often they check messages outside normal working hours. Among the 500 participants, 21.85\% always check messages outside normal working hours, whereas 20.8\% usually. 70, 110, and 107 participants never, rarely, and sometimes check messages outside normal working hour. The average Likert scale for this statement is 3.14 (\textit{i.e.}, between ``Sometimes'' and ``Usually''). Analyzing only the participants with history of mental health, 
we did not find any significant correlation for each one of the message's possible answers after applying Spearman's test. 

\textbf{Unrealistic deadlines.} The inability to meet the established deadline for a software development project completion can be problematic in many different ways. Besides the quality, economic, and business aspects, it can result in burnout since teams work overtime to try and meet the deadline. We surveyed the participants about how often they are required to meet unrealistic deadlines. 133, 172, and 121 participants never, rarely, and sometimes are required to meet unrealistic deadlines. The average Likert score for this statement is 2.34 (\textit{i.e.}, between ``Rarely'' and ``Sometimes''). Considering only the participants with history of mental health, we found an inverse correlation with participants that never are required to meet unrealistic deadlines after applying the Spearman's test (\textit{p}-value = 0.0434, \textit{rho} = -0.09).

\vspace{-0.3cm} 
\begin{mybox}{Observation 13}
\small
As more participants state that they never are required to meet unrealistic deadlines, their probability of having mental health disorders decrease.  
\end{mybox}

\textbf{COVID-19 pandemic.} Before the COVID-19 pandemic, 65.8\% of participants never worked from home. 78, 33, and 8 participants worked once a week, 2-3 times, and 4-5 times a week, respectively. The average Likert score for this statement is 1.75 (\textit{i.e.}, between ``Never'' and ``Rarely''). Considering only the participants with history of mental health, we found, applying Spearman's test, that participants that always use to work from home have a positive correlation with mental health disorders (\textit{p}-value =  0.044, \textit{rho} = 0.0898). Moreover, we found a negative correlation with participants that never use to work from home and mental health disorders (\textit{p}-value =  0.0334, \textit{rho} = -0.0951).

\vspace{-0.5cm} 
\begin{mybox}{Observation 14}
\small
Working from home before COVID-19 was a trigger to mental health disorders. Moreover, participants that never had to work from home before COVID-19 are less prone to mental health disorders.
\end{mybox}

During the COVID-19 pandemic, as expected, because of restrictions, 87.2\% of participants worked from home. Analyzing only the participants with history of mental health, we did not find any significant correlation between each one of the possible answers and participants with history of mental health disorders.

\textbf{Tasks.} We also asked the participants how often their assigned tasks change abruptly without prior notice. 76, 157, and 156 participants answered never, a few times a year, and a few times a month, respectively. The average Likert score for this statement is 2.59 (\textit{i.e.}, between ``A few times a year'' and ``A few times a month''). Analyzing the participants with history of mental health, we found an inverse correlation for participants who never had their tasks changed abruptly without prior notice (\textit{p}-value =  0.0043, \textit{rho} = -0.1272).

\vspace{-0.2cm}
\begin{mybox}{Observation 15}
\small
Participants that never have their tasks changed abruptly without prior notice are less prone to have mental health disorders.
\end{mybox}

Additionally, we asked the participants if they did not have enough time to perform their tasks with the quality they expected. 55, 137, and 125 participants strongly disagree, are neutral, and disagree, respectively, with this statement. The average Likert score for this statement is 2.96 (\textit{i.e.}, between ``Disagree'' and ``Neutral''). Analyzing the participants with history of mental health, we found an inverse correlation for participants that strongly disagree that they do not have time to do their tasks with the quality they expect (\textit{p}-value =  0.0126, \textit{rho} = -0.1113).

\vspace{-0.5cm} 
\begin{mybox}{Observation 16}
\small
Participants which agree that they have time to do their tasks with the quality they expect are less prone to have mental health disorders.
\end{mybox}

\section{Discussion}\label{discussion}

In this section, we discuss each research question in the context of the findings and survey results.

\textbf{Mental Health Disorders (RQ1) and Leisure and Entertainment activities (RQ2)}. Mental health disorders are a reality affecting software professionals. We found that 30.2\% of the survey participants were diagnosed with mental health disorders in the past or currently. Moreover, 19.4\% of participants without any record of mental health disorders were diagnosed with depression. Some participants are taking drugs without medical prescriptions and searching the web and other sources to find information about mental health. We believe that educators can use our findings and disseminate this problem and its consequences for helping practitioners since mental health is often seen as taboo in society. 

Currently, some companies are offering psychological services for employees, mainly after the COVID-19 pandemic, which affected them considerably (mental health aspects and leisure and entertainment activities). This practice must be more disseminated in the software industry, and these programs should be more proactive in identifying early symptoms of mental health disorders.

Researchers must also be involved better to understand the efficiency and limitations of these programs.  Moreover, longitudinal case studies can be designed to investigate the impact of mental health disorders in software development (quality, productivity, and well-being). While the literature discusses some aspects related to a good day for software professionals \cite{Meyer:TSE2021:SEGoodDay} and what makes them happy \cite{Graziotin:JSS2018:EffectsSEHappiness, 8851296}, little is known regarding mental health aspects. 

Finally, software professionals must engage more in leisure and entertainment activities, mainly those with a history of mental health disorders who did less physical exercise, had fewer hobbies, and spent less time with their family since these aspects can impact mental health and well-being. Researchers in future studies could also explore these aspects to investigate, for example, the software professionals' participation in these activities and the consequences for mental health (for example, monitoring the depression level of software professionals based on PHQ-9 classification during some months). 

\textbf{Work nature (RQ3)}. We did not find any correlation between organization and team size with mental health disorders. However, analyzing participants with history of mental health disorders, we found that the work type of freelancer and the role of test engineer are associated with mental health disorders. An initial hypothesis can be related to uncertainty in a contract as a freelancer, which can bring more concerns. Nevertheless, new studies must be conducted to understand this aspect and why the test engineer role is more related to mental health disorders. So far, companies and their psychological programs can better keep track of these professionals. 

\textbf{Work environment (RQ4)}. 68.8\% of participants are satisfied with the work environment, and most of the respondents (87\%) have a healthy relationship with their colleagues. Nevertheless, some aspects must be better investigated. 38\% of participants are dissatisfied with their job. Researchers and companies must work together to understand the reasons behind this aspect and the consequences for software development projects and mental health disorders. Additionally, companies and researchers must work together to investigate why software professionals do not use leisure/resting facilities and do not participate in social activities organized by the company.  

\textbf{Work schedule (RQ5)}. We identified that software professionals have worked more hours than expected, and this aspect can be a trigger to deteriorate their mental health. In addition, we found that survey participants that never are required to meet unrealistic deadlines, their tasks do not change abruptly without prior notice, and participants who agree that they have time to perform their tasks with the quality they expect are less prone to have mental health disorders. 

These aspects should be better addressed by organizations, mainly project managers responsible for project planning. We also consider that researchers should collaborate more with organizations and software professionals (managers and software engineers) to understand the current problems related to project planning (estimates, tasks size, and duration) and productivity and define new solutions to improve estimating and productivity in real settings.

\section{Limitations and Threats to Validity}\label{sec:threats}

\textbf{Internal Validity:} It is possible that the survey participants misunderstood some of the survey questions. To mitigate this threat, we employed pilot studies to test and collect feedback on our survey. We also conducted a pilot study with survey design experts. The pilot studies were also important to avoid respondent fatigue bias by making sure that the survey could be answered within 20 minutes. We updated the survey based on the findings of these pilot studies. 

\textbf{External validity:} Our findings may be limited to the survey participants. However, we believe that the large number of participants from various backgrounds more than adequately addresses this concern. 

\textbf{Construct validity}. Threats to construct validity are concerned with the extent to which the setting of the experiment reflects the construct under study, which include the potential problem of evaluation apprehension. It was mitigated by the anonymity of the participants, as well as the guaranty that all information gathered during the survey would be used only by the research team.

\section{Conclusion}\label{conclusion}

In this paper, we presented the results of the snapshot of the mental health of software professionals. To do it, we surveyed 500 software professionals from open source and private projects. 

Based on our findings, we derived 16 observations that
can be used by practitioners, organizations, and researchers. Practitioners can use our findings to maintain a healthy work-life balance in their careers, even with the specific aspects of software development. Organizations can learn from our survey respondents and take steps to remain productive while creating good programs to ensure the mental health of their employees. The research community can explore the social and human aspects to understand the impact of mental health in software development.

As future work, we intend to conduct a longitudinal case study with software professionals from a private company to gain more insights into the impact of mental health on software quality and productivity.

\bibliographystyle{IEEEtran}


\end{document}